\documentclass{bmcart}

\usepackage{amsthm,amsmath}
\usepackage{dsfont}
\usepackage{url}
\RequirePackage{natbib}
\RequirePackage{hyperref}
\usepackage{graphicx, booktabs}
\usepackage[utf8]{inputenc} %unicode support
\graphicspath{{./fig/}}

\startlocaldefs
\newcommand{\ie}{\textit{i}.\textit{e}.}
\newcommand{\eg}{\textit{e}.\textit{g}.}

\endlocaldefs

\begin{document}

%%% Start of article front matter
\begin{frontmatter}

\begin{fmbox}
\dochead{Research}

%%%%%%%%%%%%%%%%%%%%%%%%%%%%%%%%%%%%%%%%%%%%%%
%%                                          %%
%% Enter the title of your article here     %%
%%                                          %%
%%%%%%%%%%%%%%%%%%%%%%%%%%%%%%%%%%%%%%%%%%%%%%

\title{The one comparing narrative social network extraction techniques}
% The one with the extraction technique comparison
% Nodes with Friends: a comparison of social network extraction techniques
% Comparison of social network extraction techniques in narrative analysis

%%%%%%%%%%%%%%%%%%%%%%%%%%%%%%%%%%%%%%%%%%%%%%
%%                                          %%
%% Enter the authors here                   %%
%%                                          %%
%% Specify information, if available,       %%
%% in the form:                             %%
%%   <key>={<id1>,<id2>}                    %%
%%   <key>=                                 %%
%% Comment or delete the keys which are     %%
%% not used. Repeat \author command as much %%
%% as required.                             %%
%%                                          %%
%%%%%%%%%%%%%%%%%%%%%%%%%%%%%%%%%%%%%%%%%%%%%%

\author[
   addressref={aff1},                   % id's of addresses, e.g. {aff1,aff2}
   %corref={aff1},                       % id of corresponding address, if any
   %noteref={n1},                        % id's of article notes, if any
   email={michelle.edwards@adelaide.edu.au}   % email address
]{\inits{ME}\fnm{Michelle C} \snm{Edwards}}
\author[
   addressref={aff1},                   % id's of addresses, e.g. {aff1,aff2}
   %corref={aff1},                       % id of corresponding address, if any
   %noteref={n1},                        % id's of article notes, if any
   %email={lewis.mitchell@adelaide.edu.au}   % email address
]{\inits{LM}\fnm{Lewis} \snm{Mitchell}}
\author[
   addressref={aff1},                   % id's of addresses, e.g. {aff1,aff2}
   %corref={aff1},                       % id of corresponding address, if any
   %noteref={n1},                        % id's of article notes, if any
   %email={simon.tuke@adelaide.edu.au}   % email address
]{\inits{JT}\fnm{Jonathan} \snm{Tuke}}
\author[
   addressref={aff1},                   % id's of addresses, e.g. {aff1,aff2}
   %corref={aff1},                       % id of corresponding address, if any
   %noteref={n1},                        % id's of article notes, if any
   %email={matthew.roughan@adelaide.edu.au}   % email address
]{\inits{MR}\fnm{Matthew} \snm{Roughan}}

%%%%%%%%%%%%%%%%%%%%%%%%%%%%%%%%%%%%%%%%%%%%%%
%%                                          %%
%% Enter the authors' addresses here        %%
%%                                          %%
%% Repeat \address commands as much as      %%
%% required.                                %%
%%                                          %%
%%%%%%%%%%%%%%%%%%%%%%%%%%%%%%%%%%%%%%%%%%%%%%

\address[id=aff1]{%                           % unique id
  \orgname{University of Adelaide}, % university, etc
  %\street{North Terrace},                     %
  \city{Adelaide},                              % city
  \state{SA}					% state
  \postcode{5005},                                % post or zip code
  \cny{Australia}                                    % country
}

%%%%%%%%%%%%%%%%%%%%%%%%%%%%%%%%%%%%%%%%%%%%%%
%%                                          %%
%% Enter short notes here                   %%
%%                                          %%
%% Short notes will be after addresses      %%
%% on first page.                           %%
%%                                          %%
%%%%%%%%%%%%%%%%%%%%%%%%%%%%%%%%%%%%%%%%%%%%%%

\begin{artnotes}
%\note{Sample of title note}     % note to the article
%\note[id=n1]{Equal contributor} % note, connected to author
\end{artnotes}

\end{fmbox}% comment this for two column layout

%%%%%%%%%%%%%%%%%%%%%%%%%%%%%%%%%%%%%%%%%%%%%%
%%                                          %%
%% The Abstract begins here                 %%
%%                                          %%
%% Please refer to the Instructions for     %%
%% authors on http://www.biomedcentral.com  %%
%% and include the section headings         %%
%% accordingly for your article type.       %%
%%                                          %%
%%%%%%%%%%%%%%%%%%%%%%%%%%%%%%%%%%%%%%%%%%%%%%

\begin{abstractbox}

\begin{abstract} % abstract
Analysing narratives through their social networks is an expanding field in quantitative literary studies. Manually extracting a social network from any narrative can be time consuming, so automatic extraction methods of varying complexity have been developed.
However, the effect of different extraction methods on the analysis is unknown.
Here we model and compare three extraction methods for social networks in narratives: manual extraction, co-occurrence automated extraction and automated extraction using machine learning. Although the manual extraction method produces more precise results in the network analysis, it is much more time consuming and the automatic extraction methods yield comparable conclusions for density, centrality measures and edge weights. Our results provide evidence that social networks extracted automatically are reliable for many analyses. We also describe which aspects of analysis are not reliable with such a social network. We anticipate that our findings will make it easier to analyse more narratives, which help us improve our understanding of how stories are written and evolve, and how people interact with each other.
\end{abstract}

%%%%%%%%%%%%%%%%%%%%%%%%%%%%%%%%%%%%%%%%%%%%%%
%%                                          %%
%% The keywords begin here                  %%
%%                                          %%
%% Put each keyword in separate \kwd{}.     %%
%%                                          %%
%%%%%%%%%%%%%%%%%%%%%%%%%%%%%%%%%%%%%%%%%%%%%%


% MSC classifications codes, if any
%\begin{keyword}[class=AMS]
%\kwd[Primary ]{}
%\kwd{}
%\kwd[; secondary ]{}
%\end{keyword}

\end{abstractbox}
%
%\end{fmbox}% uncomment this for twcolumn layout

\end{frontmatter}

%%%%%%%%%%%%%%%%%%%%%%%%%%%%%%%%%%%%%%%%%%%%%%
%%                                          %%
%% The Main Body begins here                %%
%%                                          %%
%% Please refer to the instructions for     %%
%% authors on:                              %%
%% http://www.biomedcentral.com/info/authors%%
%% and include the section headings         %%
%% accordingly for your article type.       %%
%%                                          %%
%% See the Results and Discussion section   %%
%% for details on how to create sub-sections%%
%%                                          %%
%% use \cite{...} to cite references        %%
%%  \cite{koon} and                         %%
%%  \cite{oreg,khar,zvai,xjon,schn,pond}    %%
%%  \nocite{smith,marg,hunn,advi,koha,mouse}%%
%%                                          %%
%%%%%%%%%%%%%%%%%%%%%%%%%%%%%%%%%%%%%%%%%%%%%%

%%%%%%%%%%%%%%%%%%%%%%%%% start of article main body
% TO FIX:
% write conclusion/discussion
% table/figure captions
% edit

%%%%%%%%%%%%%%%%
%% Introduction %%
%%
\section{Introduction}
%Maybe should include related work in introduction section?
%
%\textbf{Define research territory}
%\begin{itemize}
%\item analysing narratives is good
%\end{itemize}
%
%\textbf{Narrower research territory}
%\begin{itemize}
%\item using social network analysis to understand narratives
%\end{itemize}
%
%\textbf{Identify gap}
%\begin{itemize}
%\item there are lots of different ways to extract a social network from unstructured text
%\end{itemize}
%
%\textbf{State research question}
%\begin{itemize}
%\item need to know if simpler methods to extract narrative social networks give the same analysis results as more complicated or time-consuming methods
%\end{itemize}
%
%\textbf{Approach}
%\begin{itemize}
%\item we model different methods of network extraction and simulate and compare the resulting networks
%\item as a case study we use Friends data
%\end{itemize}
%
%\textbf{Principle findings}
%\begin{itemize}
%\item we find that simpler methods have errors but overall you get the same conclusions
%\end{itemize}
%
%%%%%%%%%%%%%%%%%
%%% Background %%
%%%
%\section{Related work}
%

Quantitative narrative analysis has become increasingly popular in recent years with the availability of literary works and film and television scripts online. Narratives can be in the form of films \cite{chaturvedi-similarities, movie-emotions, neural-network}, television shows \cite{tvshows-temporal, game_of_nodes, bazzan-friends, deep-concept-hierachies} or novels \cite{les-mis, got-network, literature-social-networks, alice-network, vikingsaga, prado-literarytexts}, and reasons to analyse these include: 
\begin{itemize}
\item to gain a deeper understanding of a particular narrative, narratives of a certain type, or narratives in general, or
\item to help determine what would improve narratives in the future.
\end{itemize} 
Narrative analysis is also popular within blogs, where fans visualise data from television shows such as Game of Thrones \cite{got-family-ties}, Seinfeld \cite{seinfeld}, The Simpsons \cite{simpsons-network}, Grey's Anatomy \cite{greys-anatomy} and Friends \cite{friendsanalysis, dmathlete-friends, charts-friends}. Similarly, films \cite{starwars-network, disney-gender} and even plays \cite{shakespeare, hamilton} have been analysed quantitatively. 

Fortuin \etal\ \cite{fortuin-sequencemodels} used narrative analysis to inspire script-writers suffering from writer's block, while Gorinski \etal\ \cite{gorinski} analysed film scripts to find a logical chain of important events, allowing them to summarise film scripts automatically. We can also use narrative analysis to predict what will happen next \cite{got-network}. Event prediction in narratives also suggests potential methods for predicting real-world events from news texts \cite{granroth-eventprediction, li-constructing}.

A popular way of analysing narratives is through the social networks they describe. A social network for a narrative is comprised of characters (as nodes) and their interactions or relationships (as edges). As narratives are stories about characters' interactions \cite{les-mis}, it makes sense to analyse the narrative by analysing how the characters interact. Understanding and comparing narrative social networks could lead to insights into which structures make a narrative successful.

One of the most problematic aspects of narrative social network analysis is constructing the network from an unstructured text source such as a script or novel. Extracting an interaction network from novels is challenging because the text does not always state who is speaking. Most attempts to match quoted speech in novels to the character speaking involve Natural Language Processing (NLP) and/or machine learning techniques \cite{he-speakers, extracting-networks, agarwal_automatic, iyyer-dynamicfiction, literature-social-networks, extracting-actortopic}. A disadvantage of these techniques is that there is either significant manual work in identifying aliases of characters, or that the accuracy of character identification ranges from $<50\%$ to \(\approx 90\%\) \cite{he-speakers}. The more manual work put in at the NLP stage, the more accurate the identification tends to be. Alternatively, researchers can manually identify the speakers in novels \cite{agar-social}, but this takes substantially longer and is not practical for analysing large corpora. % could mention other ways of creating networks

Extracting social networks from film or television scripts is almost as difficult. The most accurate, but time-consuming, approach is to manually record interactions between characters \cite{bazzan-friends}. A more scalable approach is to automatically create a social network from the script of the film or television show. Scripts necessarily label speakers, but not who each character is speaking to. There are examples of using NLP and machine learning techniques \cite{chen-emotionlines, chen-identification, BoninFriends}, but again there is a trade-off with the accuracy of identifications. 

An alternative automatic method is to extract a co-occurrence network \cite{deep-concept-hierachies, got-prediction, scene-clustering, weng-movie}, which infers interactions between characters from the number of times they appear in a scene together. We can create co-occurrence networks for novels as well, for example by counting the number of times characters are mentioned within a number of words of each other \cite{got-network}. Using a co-occurrence network presumes that relationship strength can be measured by the number of times characters share a scene, as opposed to the number of times characters directly interact. While this assumption is intuitive, to the best of our knowledge, there is no research into the effect of this assumption on the resulting network properties in narrative analysis.
 
 In this paper we compare three social network extraction techniques in the context of TV scripts:
\begin{itemize}
\item manually-extracted networks (as in Bazzan \cite{bazzan-friends}),
\item networks extracted using NLP (as in Deleris \etal\ \cite{BoninFriends}), and
\item co-occurence networks extracted using scripts.
\end{itemize}
To compare these techniques we create a model to simulate interactions in a narrative. Using the simulated interactions, we create and compare observation networks based on the three extraction techniques. This \textit{in silico} model allows us to compare techniques with complete knowledge of the ground truth. Modelling the narrative also allows us to control and measure parameters such as the error rate for the NLP method or the number of scenes for the co-occurrence method. Finally, the model allows our methods to be applied to a range of narratives, not just the case study we give here.
 
We use standard network metrics (see \autoref{sec:metrics}) to compare the three different network extraction techniques, applied to the characters in the television series \textit{Friends}. \textit{Friends} is an American situation comedy (sitcom) with ten seasons aired from 1994 to 2004. We choose \textit{Friends} as a case study for our model because the series is well-known, long-running and a popular subject amongst researchers \cite{bazzan-friends, BoninFriends, chen-identification, deep-concept-hierachies, marshall-friends}.

% SUMMARY OF RESULTS
Some key findings of this work are:
\begin{itemize}
\item Co-occurrence networks have higher edge densities than the manually extracted networks, but the densities are highly correlated between techniques (the Pearson's correlation coefficient is 0.96).
\item Centrality measures (degree, betweenness, eigenvector and closeness) are highly correlated in the manually extracted networks and co-occurrence and NLP networks, but clustering is not reliable in the automated networks.
\item Edge weights in the automated networks correlate moderately with the edge weights in the manually extracted networks (the median Spearman's correlation coefficient is 0.77 for the co-occurrence networks and 0.80 for the NLP networks).
\item The six core ``friends" in \textit{Friends} (Chandler, Joey, Monica, Phoebe, Rachel and Ross) interact less with each other less as the series progresses.
\end{itemize}
We conclude that automatically extracted networks -- co-occurrence and NLP networks -- give reliable analyses for most global, character, and relationship metrics, so we recommend extracting narrative social networks in one of these ways for time efficiency. If clustering is of high importance in an analysis, however, manually extracted networks are required.

%%%%%%%%%%%%%%%%
%% Data %%
%%
\section{Data} \label{sec:data}
Although our findings are partially based on \textit{in silico} experiments, we use real data to inform our models and to provide final verification.

We examine three datasets estimating social networks for the television series \textit{Friends}. The social network describing character relationships are defined by nodes that represent characters in a chosen time frame (usually an episode or season), and edges connecting characters who interact. The precise definition of an interaction varies throughout the literature, but the assumption that characters who interact more have stronger relationships remains constant. Our goal is to model these relationships. Note that the strength of a relationship does not imply characters are good friends (despite the name of the series), as characters can have strong hostile interactions \cite{les-mis, mythological-networks}.

The first dataset consists of manually extracted data by Bazzan \cite{bazzan-friends}. It contains ordered lists of undirected interactions between pairs of characters for each episode. Bazzan manually annotated 16569 interactions from all 236 episodes, defining an interaction as two characters talking, touching or having eye contact. While there may be human interpretation errors in this dataset, this is the most reliable method of extracting the social network. Therefore, the manual extraction method provides a `gold standard' for the social networks of the characters. We call the networks from this dataset the manually extracted networks, or \textbf{manual networks}. The edge weights in the \textbf{manual networks} correspond to the number of interactions between two characters in a given timeframe. \autoref{tab:data} shows the number of episodes, interactions, scenes and characters in each season.

% latex table generated in R 3.5.1 by xtable 1.8-2 package
% Mon Oct 22 12:37:53 2018
\begin{table}[ht]
\centering
\begin{tabular}{r|rrrrrrr}
  \toprule
Season & Episodes & Chars & Ints & Scenes & Ints/Episode & Scenes/Episode & Ints/Scene \\ 
  \midrule
  1 &  24 & 126 & 2492 & 364 & 103.83 & 15.17 & 6.85 \\ 
    2 &  24 & 107 & 1815 & 314 & 75.62 & 13.08 & 5.78 \\ 
    3 &  25 &  98 & 1770 & 422 & 70.80 & 16.88 & 4.19 \\ 
    4 &  24 &  96 & 1598 & 438 & 66.58 & 18.25 & 3.65 \\ 
    5 &  24 &  92 & 1786 & 378 & 74.42 & 15.75 & 4.72 \\ 
    6 &  25 &  99 & 1491 & 387 & 59.64 & 15.48 & 3.85 \\ 
    7 &  24 &  81 & 1475 & 402 & 61.46 & 16.75 & 3.67 \\ 
    8 &  24 & 110 & 1220 & 356 & 50.83 & 14.83 & 3.43 \\ 
    9 &  24 & 101 & 1454 & 345 & 60.58 & 14.38 & 4.21 \\ 
   10 &  18 &  88 & 1468 & 238 & 81.56 & 13.22 & 6.17 \\ 
   \bottomrule
\end{tabular}
\caption{Summary of data from manually collected dataset \cite{bazzan-friends}. For each season we have the number of episodes (Episodes), the number of characters (Chars), the total number of interactions (Ints), and the number of scenes (Scenes). We also calulate the number of interactions per episode (Ints/Episode), number of scenes per episode (Scenes/Episode) and average number of interactions per scene for each season (Ints/Scene).} 
\label{tab:data}
\end{table}

\autoref{tab:data} shows there are 24 episodes in most seasons, but 25 episodes in Season 3 and Season 6 and only 18 episodes in Season 10. Season 1 has notably more interactions than any other season, possibly due to the need to establish characters and relationships at the beginning of the series. We will discuss our findings on trends in network properties over all 10 seasons in \autoref{sec:lessfriendly}.

The second dataset contains co-occurrence networks, extracted using scripts available from a fan website \cite{friends-scripts}. Using Python \cite{python} we processed the scripts, and identified scene breaks and the characters that speak in each scene. For every scene, we assume every character interacts with every other character, so a scene is a ``co-occurrence", and hence we call these networks \textbf{co-occurrence networks}. The edge weights correspond to the number of co-occurrences between characters. 

Note that only speaking characters are identified, even though there could be scenarios where characters appear or interact without speaking. Also, the scripts \cite{friends-scripts} were transcribed manually, so there are issues with inconsistency, typing mistakes and characters being referred to by different names (\eg\ Ross, Ross Geller or Mr. Geller). To minimise these issues, we cleaned the scripts using manually defined regular expressions. The resulting \textbf{co-occurrence networks} dataset contains weighted edge lists for 227 episodes. The total number of episodes is less than for the \textbf{manual networks} because episodes with two parts (\eg\ S1E16 - The One With Two Parts: Part One and S1E17 - The One With Two Parts: Part Two) are included as one episode in the \textbf{co-occurrence networks} dataset. The total number of interactions over all \textbf{co-occurrence network} episodes is 18574, which is surprisingly close to that of the \textbf{manual networks} dataset.

An NLP network dataset for \textit{Friends} was not available, but Deleris \etal\ \cite{BoninFriends} provide information about how they extracted the social network, making use of Chen and Choi's data \cite{chen-identification}. Chen and Choi use NLP techniques to identify which character is mentioned when another character says `you', `he', `they', etc. They estimate their model correctly identifies a character 69.21\% of the time. Deleris \etal\ use `character mention' information to build a directed social network where the interactions are one of four kinds of signals: 
\begin{itemize}
\item Direct Speech (\eg\ A talks to B).
\item Direct Reference (\eg\ A says `I like you' to B).
\item Indirect Reference (\eg\ A says `I like B').
\item Third-Party Reference (\eg\ C says `A likes B').
\end{itemize}
Each of these is an example of a directed interaction from A to B. However, for the purpose of modelling networks consistently between all three approaches, we assume all interactions are reciprocated. We call the undirected networks extracted using this approach the \textbf{NLP networks}.

%%%%%%%%%%%%%%%%
%% Method %%
%%
\section{Method}

\subsection{Overview}

% FIGURE fig:method
\begin{figure}
\includegraphics[width = 0.83\textwidth]{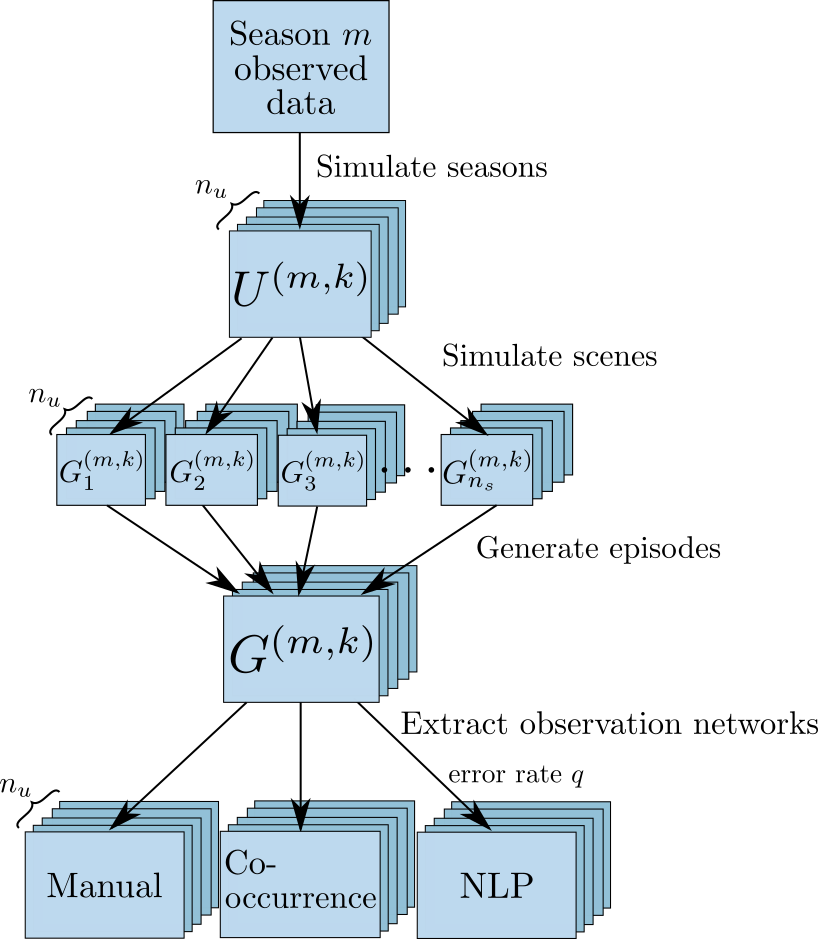}
\caption{Flow chart describing the simulation process. We model each season \(m\) in the observed data to simulate underlying season networks \(U^{(m,k)}\), where \(k=1,\ldots,n_{u}\) (\autoref{sec:simseason}). From each underlying season network we simulate scene networks \(G_{\ell}^{(m,k)}\) for \(\ell=1,\ldots,n_s\) (\autoref{sec:simscene}). The scene networks combine to generate episode networks \(G^{(m,k)}\) (\autoref{sec:simepisode}). From each simulated episode network we extract three observation networks; a \textbf{manual network}, a \textbf{co-occurrence network} and an \textbf{NLP network} (\autoref{sec:simobservations}).}
\label{fig:method}
\end{figure}

We compare the network extraction methods by simulating narrative social networks, `extracting' observed networks using the three extraction methods and comparing these observed networks. We simulate social networks using a data-driven model. Simulation allows us to generalise the problem to any narrative that has a similar underlying social network and to generate large datasets for statistical analyses. The simulation and extraction process is outlined in \autoref{fig:method} and the following sections.

We estimate parameters for our model using the \textbf{manual networks} data, then use the model to simulate \(n_u\) underlying season networks. For each season network, we use a random walk process to simulate \(n_s\) scenes. We then combine the scenes to form a simulated episode. From each simulated episode, we extract three observation networks resembling the \textbf{manual networks}, \textbf{co-occurrence networks} and \textbf{NLP networks}. We compare these simulated networks using the network metrics outlined in \autoref{sec:metrics}.

\subsection{Simulate season from data} \label{sec:simseason}
The first step described by \autoref{fig:method} is simulating underlying season networks from the observed data. To simulate networks we need to model the seasons in the \textbf{manual networks} dataset. We want, in addition to edges, to simulate edge weights, non-negative integers representing the number of character interactions. We notice there are significant differences between the way the core characters of \textit{Friends} (Monica, Rachel, Phoebe, Ross, Chandler and Joey) interact with each other (average of 81 interactions per pair per season) and with other characters (average of 0.71 interactions per pair per season), and the way other characters interact with each other (average of 0.0093 interactions per pair per season). We therefore propose a two-class Poisson model for each season of the \textbf{manual networks}. 

Let \(V^{(m)}= \left\{1,\ldots,N^{(m)}\right\}\) be the set of characters in Season \(m\) and \(w_{ij}^{(m)}\geq0\) be the number of interactions between character \(i\) and character \(j\) in Season \(m\) of the \textbf{manual networks} dataset. We partition \(V^{(m)}\) such that
\[V^{(m)} = V_{\text{core}}\cup V_{\text{non-core}}^{(m)},\]
where \(V_{\text{core}}\) contains the 6 core characters who are constant across all seasons, and \(V_{\text{non-core}}^{(m)}\) contains the \(\left(N^{(m)}-6\right)\) non-core characters for Season \(m\).

For the two-class Poisson model, assume each edge weight \(w_{ij}^{(m)}\) in Season \(m\) of the \textbf{manual networks} dataset is a random observation of 
\[W_{ij}^{(m)} \sim \text{Poi}\left(\lambda_{C_i C_j}^{(m)}\right),\]
where
\[C_i =
\begin{cases}
1 & \text{if } i\in V_{\text{core}},\\
0 & \text{if } i\in V_{\text{non-core}}.
\end{cases}
\]

We estimate \(\lambda_{C_i C_j}^{(m)}\) using maximum likelihood estimation with the \textbf{manual network} edge weights:
\[
\hat{\lambda}_{C_iC_j}^{(m)} =
\begin{cases}
\displaystyle\frac{\sum_{i<j} w_{ij}^{(m)} C_i C_j}{\sum_{i<j} C_i C_j} & \text{if }C_i C_j = 1,\\
\displaystyle\frac{\sum_{i<j} w_{ij}^{(m)} (1-C_i)(1-C_j)}{\sum_{i<j} (1-C_i)(1-C_j)} & \text{if }(1-C_i) (1-C_j) = 1,\\
\displaystyle\frac{\sum_{i<j} w_{ij}^{(m)} (C_i + C_j - 2C_iC_j)}{\sum_{i<j} (C_i + C_j - 2C_iC_j)} & \text{otherwise.}
\end{cases}
\]

For each season \(m\) we simulate \(n_u\) season networks
\[U^{(m,k)} = \left(V^{(m)}, E^{(m,k)}\right),\]
where \(k=1,\ldots,n_u\) and 
\[E^{(m,k)} = \left\{W_{ij}^{(m,k)} \mid i,j \in V^{(m)}, i<j \right\}.\]
Note that each simulation contains all characters \(V^{(m)}\) from Season \(m\), but the edge weights are randomised. We generate random edge weights between each pair of nodes from the distribution
\[W_{ij}^{(m,k)} \sim \text{Poi}\left( \hat{\lambda}_{C_iC_j}^{(m)} \right).\]
This method allows for edges with zero-weights. We take zero-weights to mean there are no interactions between the characters, which is equivalent to having no edge between the characters.

\subsection{Simulate scene from season} \label{sec:simscene}

Given an underlying season network \(U^{(m,k)}\), we wish to sample an episode network, every episode being a sequence of scenes. Following Fortuin \etal\ \cite{fortuin-sequencemodels}, we define a scene as a story part with a constant set of characters in a constant location. This approximation allows a consistent comparison between methods. Each scene also contains a set of interactions, so we can form a social network for every scene. Interactions within a scene are dependent. For example, if Joey talks to Monica, it is likely that Monica will then talk to Joey. We capture this in the model by proposing a random walk model for interactions in each scene.

The random walk model randomly picks a starting character in \(V^{(m)}\), with probability proportional to the eigenvector centrality of the character in \(U^{(m,k)}\) (see \autoref{sec:metrics}). This character randomly interacts with another character with probability proportional to the edge weight between the characters. That character randomly interacts with another character, selected in the same way. The random walk process continues until we reach \(n_{\text{int}, \ell}\) interactions. We choose \(n_{\text{int}, \ell}\) based on the average number of interactions per scene in the data from \autoref{tab:data}. The next scene starts with a new random starting character so that we have a fresh set of characters in each scene.

Each scene \(\ell\) consists of a set of characters \(C_{\ell}^{(m,k)}\) and interactions \(L_{\ell}^{(m,k)}\). Let \(n_s^{(m)}\) be the rounded average number of scenes per episode in Season \(m\) from our datasets. We define the network of a scene \(\ell\) sampled from \(U^{(m,k)}\) as
\[G_{\ell}^{(m,k)} = \left(C_{\ell}^{(m,k)}, L_{\ell}^{(m,k)}\right),\]
where \(\ell = 1,\ldots, n_s^{(m)}\). As shown in \autoref{fig:method}, we independently simulate \(n_s\) scenes with the random walk model from each simulated season network \(U^{(m,k)}\), then combine the scenes to generate a random episode.

\subsection{Generate episode from simulated scenes} \label{sec:simepisode}

We generate an episode by concatenating simulated scenes. An episode sampled from \(U^{(m,k)}\) is 
\[G^{(m,k)} = \left(G_{1}^{(m,k)}, G_{2}^{(m,k)}, \ldots, G_{n_s}^{(m,k)}\right).\]
The set of characters in \(G^{(m,k)}\) are the union of the sets of characters in the scenes, \ie\ \(\bigcup_{\ell=1}^{n_s} C_{\ell}^{(m,k)}\). The edge weight between character \(i\) and \(j\) in \(G^{(m,k)}\) is the sum of the interactions between characters \(i\) and \(j\) in the scene networks, which is zero if at least one of \(i\) or \(j\) was not in the scene.

\subsection{Extract observation networks from simulated episodes} \label{sec:simobservations}

As in \autoref{fig:method}, we extract three observed networks from each simulated episode \(G^{(m,k)}\); a \textbf{manual network}, a \textbf{co-occurrence network} and an \textbf{NLP network}. We compare these simulated networks using metrics outlined in \autoref{sec:metrics}.

The \textbf{manual network} is built from the actual data so it is assumed to be 100\% correct. Therefore the simulated \textbf{manual network} extracted from \(G^{(m,k)}\) is \(G^{(m,k)}\).

The \textbf{co-occurrence network} is obtained by creating a clique for the characters in each scene. We add clique networks so that edge weights correspond to the number of scenes two characters are in together, as they would be in the automated process.

The \textbf{NLP network} counts interactions similarly to the \textbf{manual network}, however it simulates NLP by including errors in the identification of characters. We model these errors by assuming:
\begin{enumerate}
\item One character (the speaking character) has been identified correctly, but the character being spoken to may be misidentified with probability \(q\).
\item An incorrectly identified character is equally likely to be any character in the episode except the speaking character or correct character.
\end{enumerate}
Chen and Choi \cite{chen-identification} obtained a ``purity score" of 69.21\% in their analysis of \textit{Friends}, which they describe as the effective accuracy of character identification and hence we set \(q=0.3\). The impact of \(q\) as it changes is a potential direction for future work. We call the process of incorrect character identification ``rewiring".

In practice, the definition of an interaction (and hence edge weight) differs in \textbf{NLP networks} compared to \textbf{manual networks}. In the \textbf{manual networks} an interaction occurs between two characters who see, talk to or touch each other, whereas an interaction in the \textbf{NLP networks} occurs when two characters talk to, mention or refer to each other. We do not have this information in the \textbf{manual networks}, so we assume that characters seeing and touching each other is equivalent to characters mentioning and referring to each other.

\subsection{Comparing observation networks}\label{sec:metrics}

To measure how social network extraction method affects narrative analysis we compare the simulated \textbf{manual networks} with the simulated \textbf{co-occurrence} and \textbf{NLP networks}, using three types of network metrics;
\begin{enumerate}
\item \emph{Global metrics}: size, total edge weight, edge density and clustering coefficient.
\item \emph{Node/character metrics}: degree, betweenness centrality, eigenvector centrality, closeness centrality and local clustering coefficient.
\item \emph{Edge/relationship metrics}: edge weights.
\end{enumerate}
These metrics are common in narrative social network analysis, providing useful insight into social structure and important characters and relationships. The aim is not to compare the observation networks directly, but to investigate the effect the different observation types have on the narrative analysis. Consequently, we are more interested in understanding how metrics correlate rather than systematic differences in their value.

Let \(G = (V, E)\) be an observed network with nodes \(V = \{i\in \mathds{N}\mid 1\leq i \leq n\}\) and weighted edges \(E = \{w_{ij} \in \mathds{N}| i,j\in V\}\). Note that our networks are undirected and loop-free so \(w_{ij} = w_{ji}\) and \(w_{ii}=0\) \(\forall i,j\in V\). We define a geodesic between node \(i\) and node \(j\) by the path containing the least number of edges, ignoring edge weights. We can then define the following commonly used metrics for narrative social network analysis.

\subsubsection{Global metrics}
\begin{itemize}

\item The \textit{size} of \(G\) is the number of nodes/characters, \(|V|\).

\item The \textit{total edge weight} of \(G\) is the sum of all edge weights, or number of interactions.

\item The \textit{density} of \(G\) is the proportion of observed edges.

\item The \textit{clustering coefficient} of \(G\) is
\[C(G) = \frac{\text{(number of triangles)}\times 3}{\text{(number of connected triples)}},\]
where a triangles is three vertices with three edges and a connected triple is three vertices with at least two edges, as in Newman \cite{network-book}, page 183.

\end{itemize}

\subsubsection{Character metrics}
\begin{itemize}

\item The \textit{weighted degree} of node \(i\) is the sum of the weights of the edges adjacent to node \(i\). This is the number of times character \(i\) interacts.

\item The \textit{betweenness centrality} of node \(i\) is
\[between(i) = \sum_{s,t}\frac{n_{st}^i}{g_{st}},\]
where \(n_{st}^i\) is the number of geodesics from node \(s\) to \(g\) that go through node \(i\) and \(g_{st}\) is the total number of geodesics from node \(s\) to \(t\) (see Newman \cite{network-book}, page 173). \textit{Betweenness centrality} measures the extent to which a character connects characters to other characters.

\item The \textit{eigenvector centrality} of node \(i\) is the \(i\)th element of the eigenvector corresponding to the largest eigenvalue of the adjacency matrix of G (see Newman \cite{network-book}, page 159). \textit{Eigenvector centrality} measures the influence of a character in the network by weighting connections with more highly connected characters more highly.

\item The \textit{closeness centrality} of node \(i\) is the inverse of the average length of geodesics from node \(i\):
\[close(i) = \frac{1}{\sum_j d_{ij} / n},\]
where \(d_{ij}\) is the number of edges in the geodesic between node \(i\) and node \(j\) (see Newman \cite{network-book}, page 170). \textit{Closeness centrality} measures how close a character is to all the other characters, which can indicate whether the character is central to the narrative.

\item The \textit{local clustering coefficient} of node \(i\) is the proportion of triangles centred on node \(i\) that are closed (see Newman \cite{network-book}, page 186). It measures the degree to which node \(i\) is part of a cluster.

\end{itemize}

\subsubsection{Relationship metrics}
\begin{itemize}

\item The \textit{edge weight} of the edge between node \(i\) and node \(j\) is \(w_{ij}\). In our narrative social network context \(w_{ij}\) represents the number of interactions that occur between character \(i\) and character \(j\).

\end{itemize}

%%%%%%%%%%%%%%%%
%% Results %%
%%
\section{Results}
\begin{figure}[h]
\includegraphics[width = \textwidth]{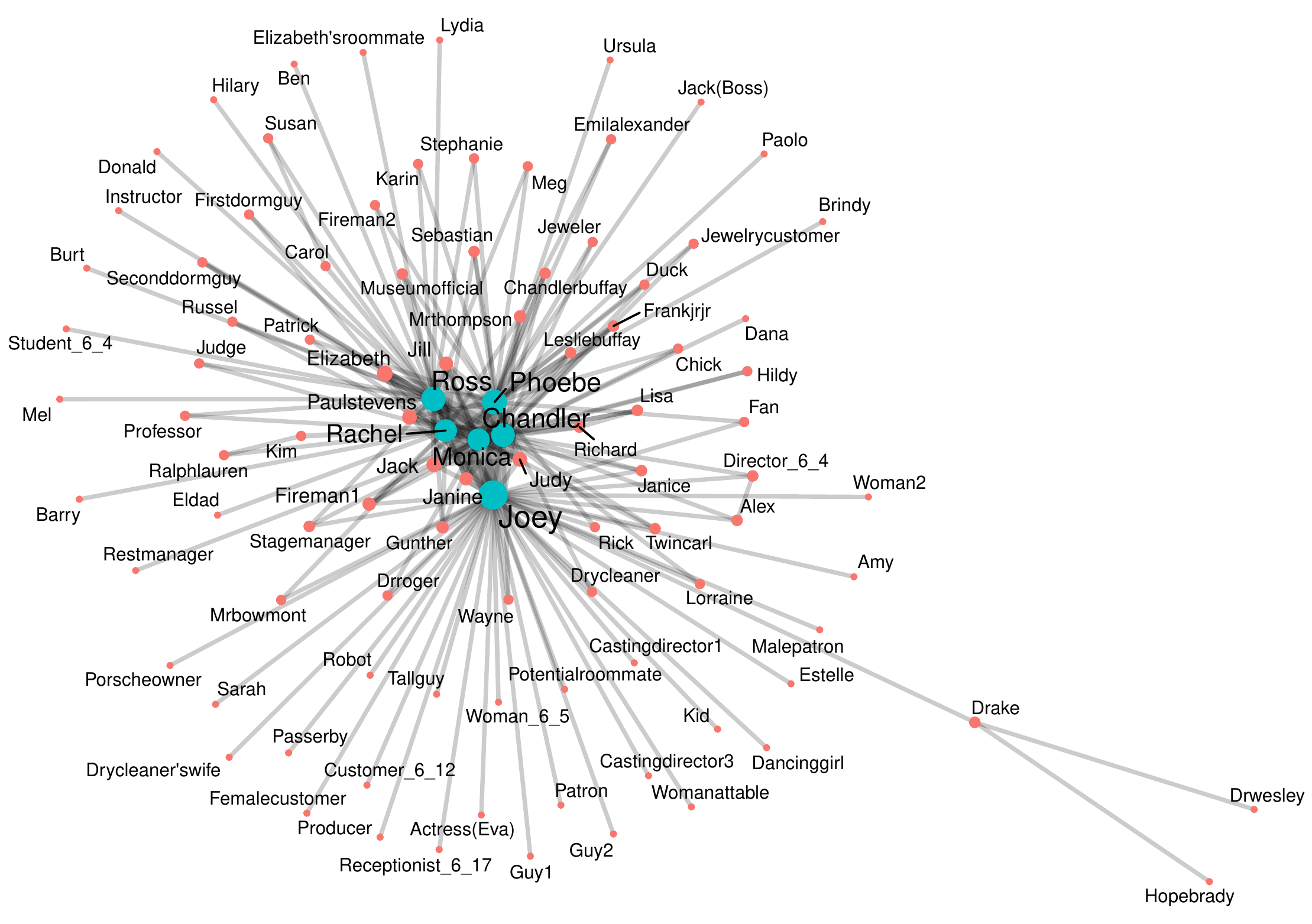}
\caption{Network of Season 6 from the manually extracted network dataset. The core characters have blue nodes and other characters have red nodes. The width of the edges represent the edge weight and the size of the nodes represent the node degree.}
\label{fig:season6}
\end{figure}

We simulate \(n_u=10000\) seasons using the two-class Poisson model on Season \(m=6\), as the number of scenes per episode in Season 6 is close to the mean number of scenes per episode over all ten seasons. \autoref{fig:season6} shows the social network of interactions from Season 6. From each simulation, we sample interactions from one episode using random walks for each scene. \autoref{tab:data} shows Season 6 of Friends has 25 episodes, 1491 interactions and 379 scenes. Therefore the average episode in season 6 has approximately 60 interactions and 20 scenes. We set \(n_s = 15\) scenes and \(n_{int,\ell} = 4\) for every scene \(\ell\).
% FIGURE fig:season5

From each sample episode network we `extract' the three observation networks using the methods described in \autoref{sec:simobservations} and compare using the metrics described in \autoref{sec:metrics}. We find that there are differences in the value of the metrics across observation networks, but the errors are mostly systematic. While the exact values of metrics can vary across the different observation networks, the important features in the narrative analysis (\ie\ the rankings and trends of metrics) would not be greatly affected. The global metrics of the simulated \textbf{co-occurrence networks} and \textbf{NLP networks} correlate to those of the associated \textbf{manual networks}. The centrality metrics (degree, betweenness, eigenvector and closeness centrality) also have high correlation with the same character metrics across the simulated \textbf{manual} and random observation networks, but there is wide variance in the correlations of the local clustering coefficient of characters. The edge weights in the simulated \textbf{manual networks} also correlate reasonably highly with the edge weights in the simulated \textbf{co-occurrence} and \textbf{NLP networks}. 

\subsection{Global metric comparison}

\begin{figure}[h]
\includegraphics[width = \textwidth]{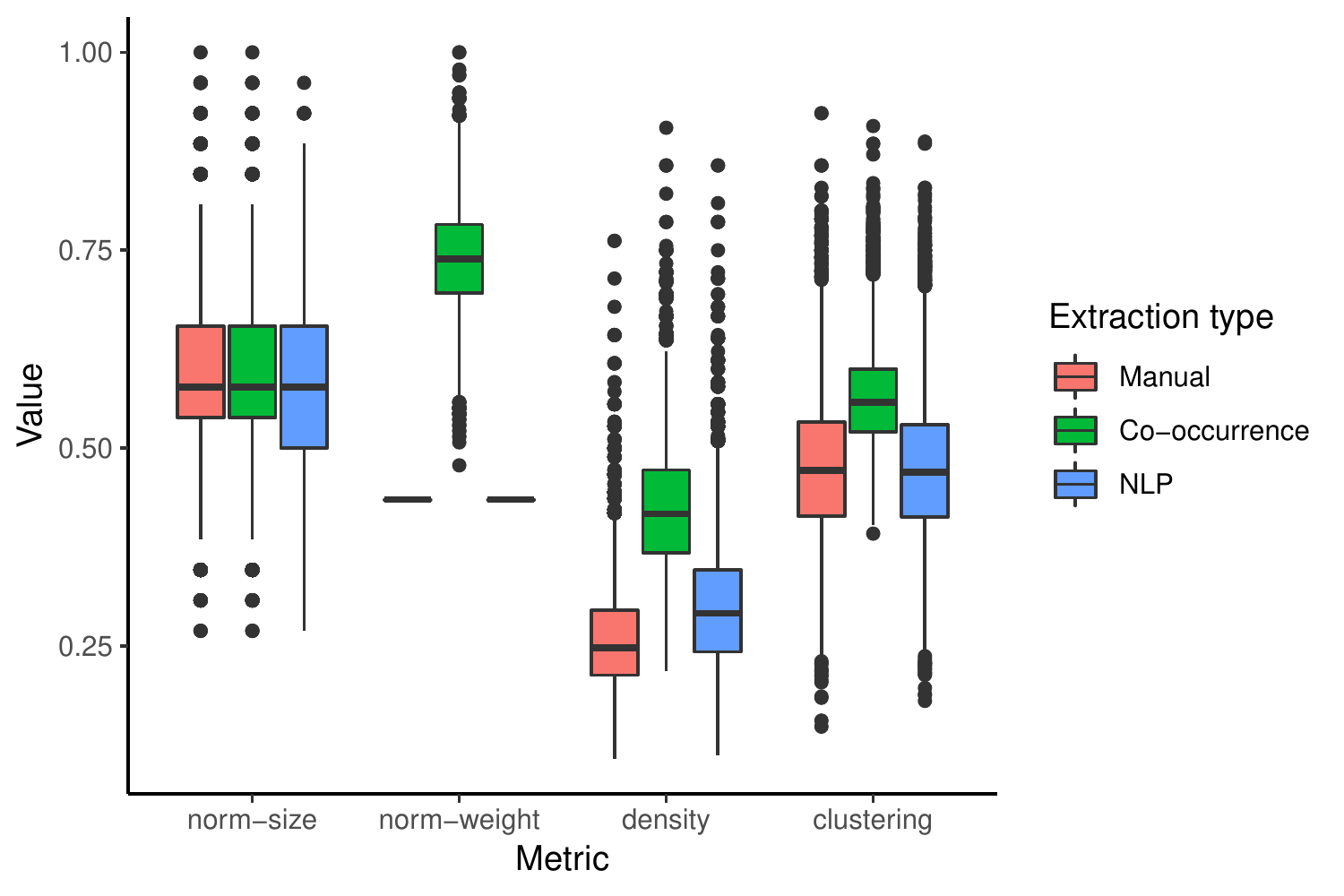}
\caption{Box-plots of the normalised size, normalised total edge weight, edge density and clustering coefficients of the \textbf{manual}, \textbf{co-occurrence} and \textbf{NLP networks} from the 10000 simulations. The size and total edge weights are normalised by dividing by the maximum.}
\label{fig:global}
\end{figure}

\autoref{fig:global} show boxplots of the normalised size, total edge weight, density and clustering for each network type. We normalise size and total edge by dividing by the maximum over the three network types.

% FIGURE fig:global

The simulated \textbf{manual} and \textbf{co-occurrence networks} have the same number of characters (and hence size) in each simulation. However, it is possible in real data to see differences. We find that the discrepancies in size of the real datasets are almost always due to differences in what defines a character. For example, `answering machine' is counted as a character in one \textbf{co-occurrence network}, but not in the corresponding \textbf{manual network}. The size of the simulated \textbf{NLP network} is always equal to or less than the size of the other simulated observation networks, as our model can only rewire to characters within the true episode. Characters are excluded if all the edges connected to that character are rewired away and no edges are rewired back to the character. This is more likely to happen to characters that are connected to few edges in the first place, and so the effect of the rewiring on the analysis is minimal.

The simulated \textbf{manual} and \textbf{NLP networks} have the same number of interactions (and hence total edge weight) in each simulation by construction. In practice there might be discrepancies in total edge weights due to different definitions of interactions as discussed in \autoref{sec:data} and \autoref{sec:simobservations}. The simulated \textbf{co-occurrence networks} have between 10\% and 100\% more interactions on average. We see this in \autoref{fig:global} through an increase in the normalised weight and edge density.

Interestingly, it is rare for the simulated \textbf{NLP network} to have a lower edge density than the simulated \textbf{manual network} even though the edges are rewired with equal probability to any other character. This occurs because we only rewire one interaction, not the entire edge with its weight.

% FIGURE
\begin{figure}[h]
\includegraphics[width = 0.8\textwidth]{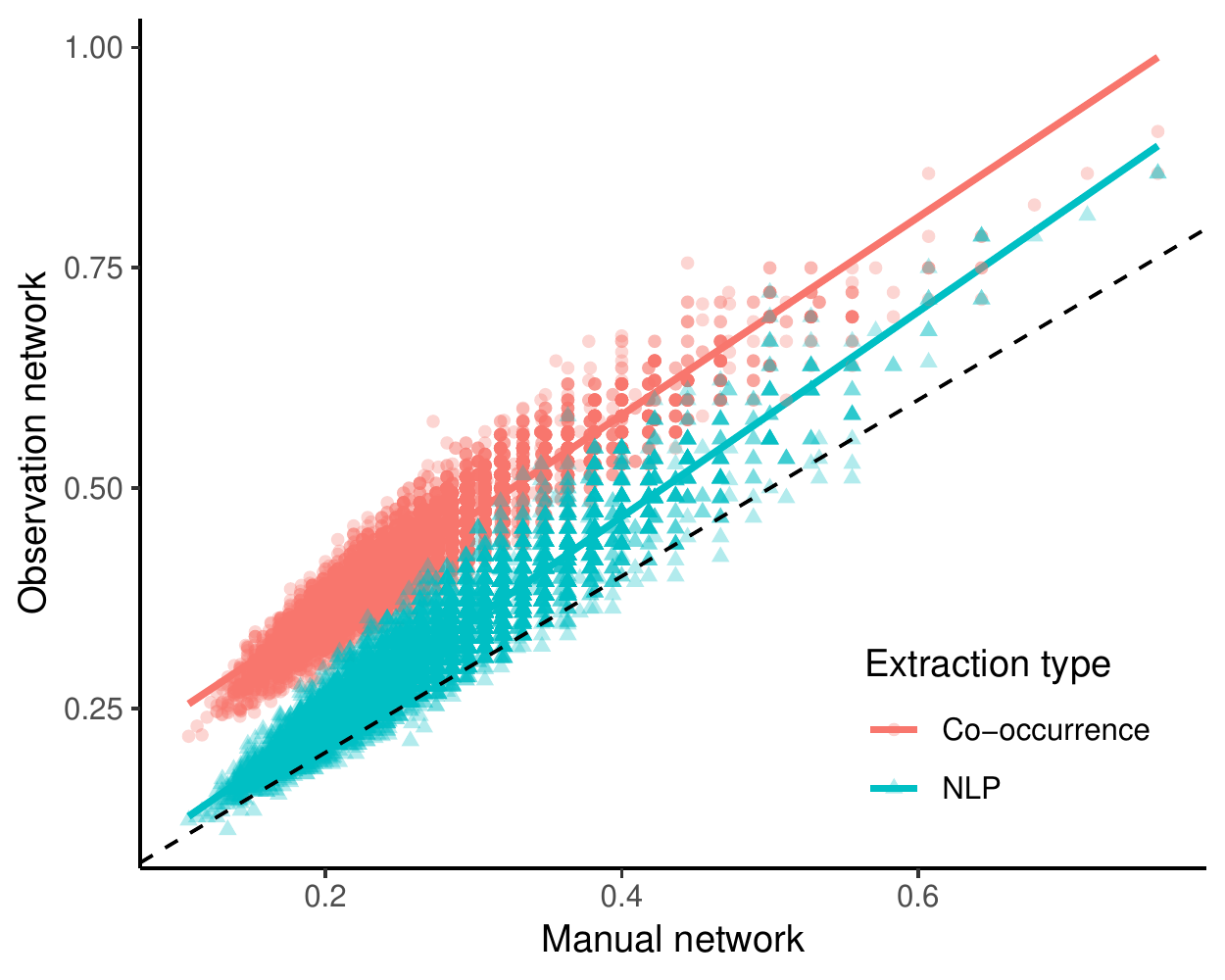}
\caption{Edge density of the simulated \textbf{manual network} compared to the simulated \textbf{co-occurrence} and \textbf{NLP networks}.  The dashed line shows \(y=x\). The edge densities of the simulated \textbf{co-occurrence network} are much larger than those of the \textbf{manual network} because creating cliques every scene adds many edges. The simulated \textbf{NLP network} edge densities are usually larger than those of the \textbf{manual network} because of the rewiring process. The edge densities of both random observation networks are linearly related to the simulated \textbf{manual network} edge densities.}
\label{fig:density}
\end{figure}

However, there is very high correlation between the \textbf{manual network} density and the other observation networks. \autoref{fig:density} shows a scatterplot of the two. The Pearson correlation coefficient between the simulated \textbf{manual} and \textbf{co-occurrence network} edge densities is 0.96, and between the simulated \textbf{manual} and \textbf{NLP network} edge densities is 0.95. This means that while there is some systematic bias, comparing social networks using relative edge density is not greatly affected. Importantly, the different extraction methods don't distort trends.

\begin{figure}[h]
\includegraphics[width = 0.8\textwidth]{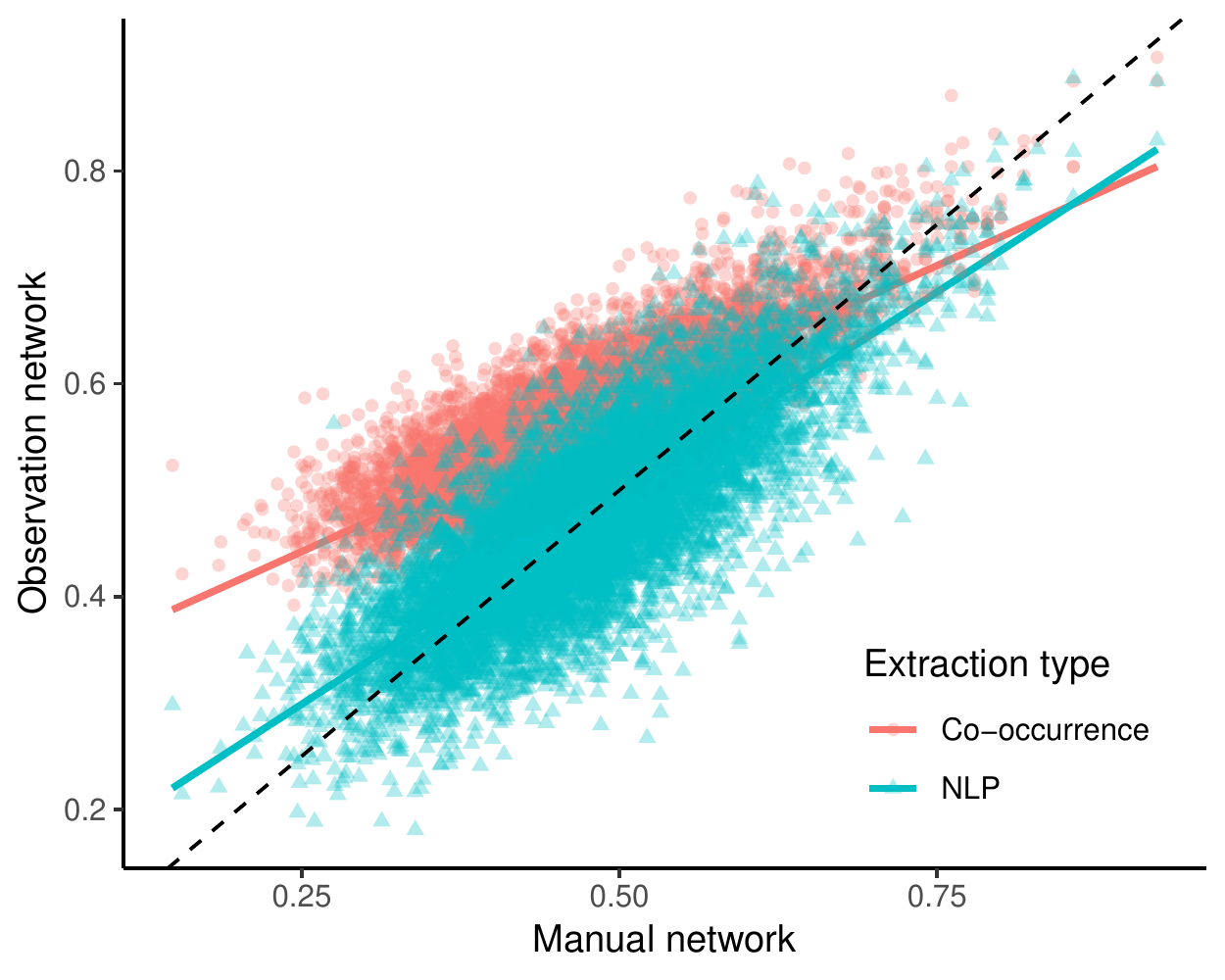}
\caption{Clustering coefficient of the simulated \textbf{manual network} containing all interactions compared to the simulated \textbf{co-occurrence} and simulated \textbf{NLP networks}. The dashed line shows \(y=x\). The clustering coefficients of the simulated \textbf{co-occurrence networks} are larger than of the \textbf{manual networks} because we create cliques in every scene. The clustering of the simulated \textbf{NLP networks} is similar to the clustering of the simulated \textbf{manual networks} but there is variation.}
\label{fig:clustering}
\end{figure}

\autoref{fig:global} shows that the simulated \textbf{co-occurrence networks} are more clustered than the simulated \textbf{manual networks}. This is not surprising as forming cliques for every scene creates clusters. \autoref{fig:clustering} shows a scatterplot of the relationships, showing that the increase in clustering from the simulated \textbf{manual} to \textbf{co-occurrence network} is smaller for more highly clustered networks. This occurs because if the \textbf{manual network} is already highly clustered, forming cliques in every scene will add fewer interactions between characters. Unclustered networks, however, will appear clustered using the co-occurrence method, so analysis of clustering is not reliable in \textbf{co-occurrence networks}. This is the largest non-systematic distortion we see accross the different techniques.

The clustering coefficients of the simulated \textbf{NLP networks} are similar to those of the simulated \textbf{manual networks} (Pearson's correlation coefficient of 0.80), but there is some variation due to rewiring interactions. Therefore, when analysing clustering in the networks, \textbf{NLP networks} are reliable in general.

% FIGURE

% real data comparison
%Our results for the global metrics for the manual observation networks and co-occurrence observation networks are consistent to what we observe in the datasets. The size of the networks are generally the same, with a few discrepancies as discussed. There are more interactions in the co-occurrence networks than the manually extracted networks, but the differences are smaller than the differences in the model. In the co-occurrence dataset there are sometimes more components than in the manually extracted networks, but these are due to errors in identifying scenes which was not taken into account in the model. The edge densities and clustering coefficients in the co-occurrence networks are also greater than those in the manually extracted networks.

\subsection{Character metrics}

\begin{figure}[h]
\includegraphics[width = 0.9\textwidth]{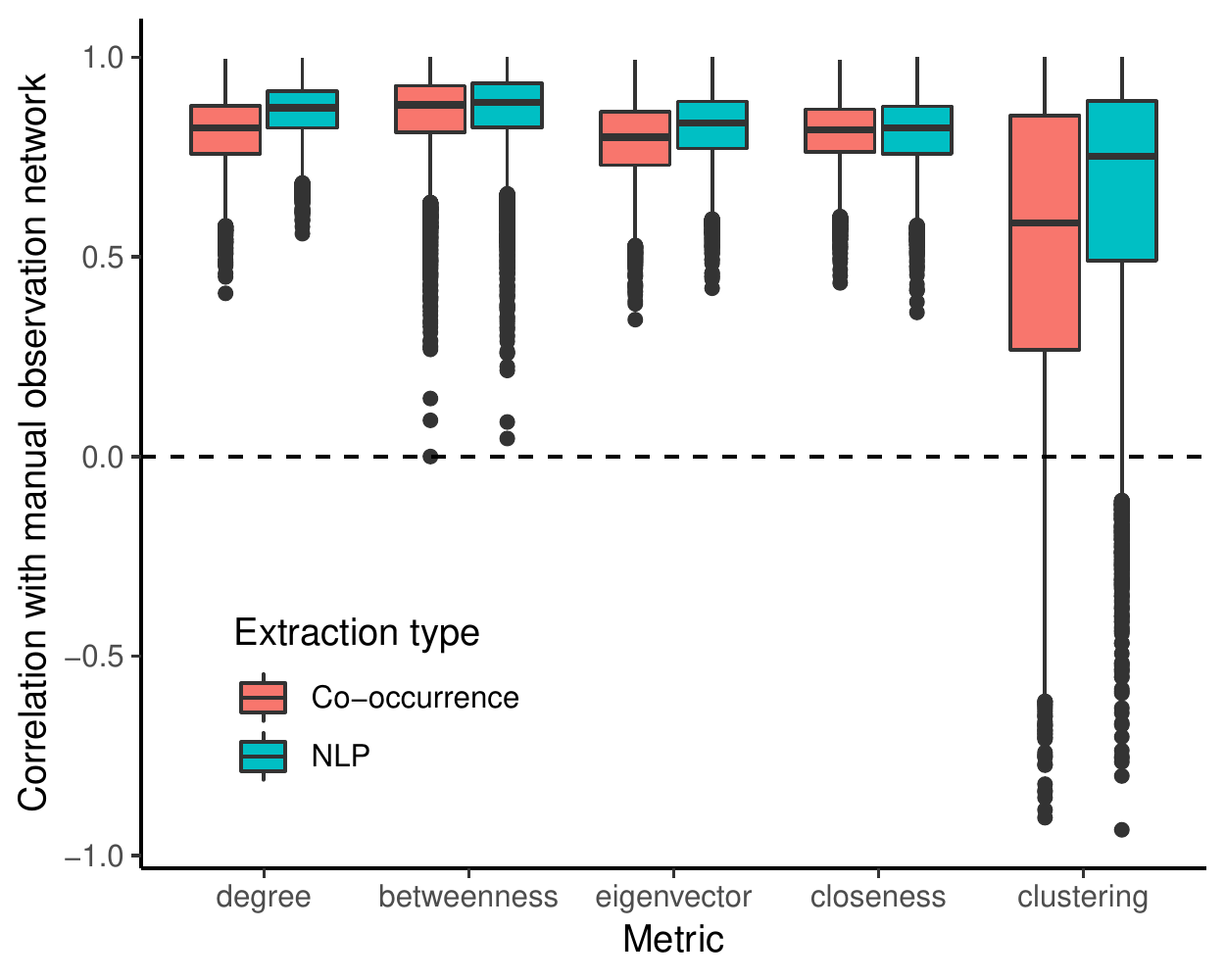}
\caption{Box-plots of the character metric correlations between the simulated \textbf{manual network} and simulated \textbf{co-occurrence} and \textbf{NLP networks }from the 10000 simulations. The degree, eigenvector and closeness centrality correlations are high with little variance. The betweenness centrality correlations are reasonably high but with more outliers. The clustering correlations have a large range and are extremely variable.}
\label{fig:character}
\end{figure}

Node/character metrics are used in narrative social network analysis to investigate the role of each character. Generally narratives are made up of a variety of characters with different roles. For example, Agarwal \etal\ \cite{agar-social} determined that Alice was the main storyteller in Lewis Carroll's \textit{Alice in Wonderland}, whereas Mouse's main role was to introduce other characters to Alice. Similarly Bazzan \cite{bazzan-friends} showed that in \textit{Friends}, while Joey connects many characters, Monica interacts the most with the five other main characters. Degree, betweenness centrality, eigenvector centrality, closeness centrality and local clustering coefficient are commonly used to assess the relative importance of the characters. We care more about comparisons between characters \eg\ who is the most central, so we examine the correlation between character metrics in the different networks, not the actual values. We use Spearman's correlation coefficient because we are interested in the rankings of importance of characters. \autoref{fig:character} shows box-plots of these correlations for weighted degree, betweenness, eigenvector and closeness centrality, and local clustering coefficient.

% FIGURE fig:character

The network observation type affects the centrality metrics (weighted degree, betweenness, eigenvector and closeness) similarly. Correlations between centrality rankings of characters in both simulated \textbf{co-occurrence} and \textbf{NLP networks} and simulated \textbf{manual networks} are high (approximately 0.8) on average, and there is little variance. This suggests that observation type does not have a strong effect on character ranking by centrality score. Therefore we can be confident in the results when analysing a narrative through the centrality of its characters in a \textbf{co-occurrence network} or \textbf{NLP network} .

% FIGURE fig:eigen
\begin{figure}[h]
\includegraphics[width = 0.9\textwidth]{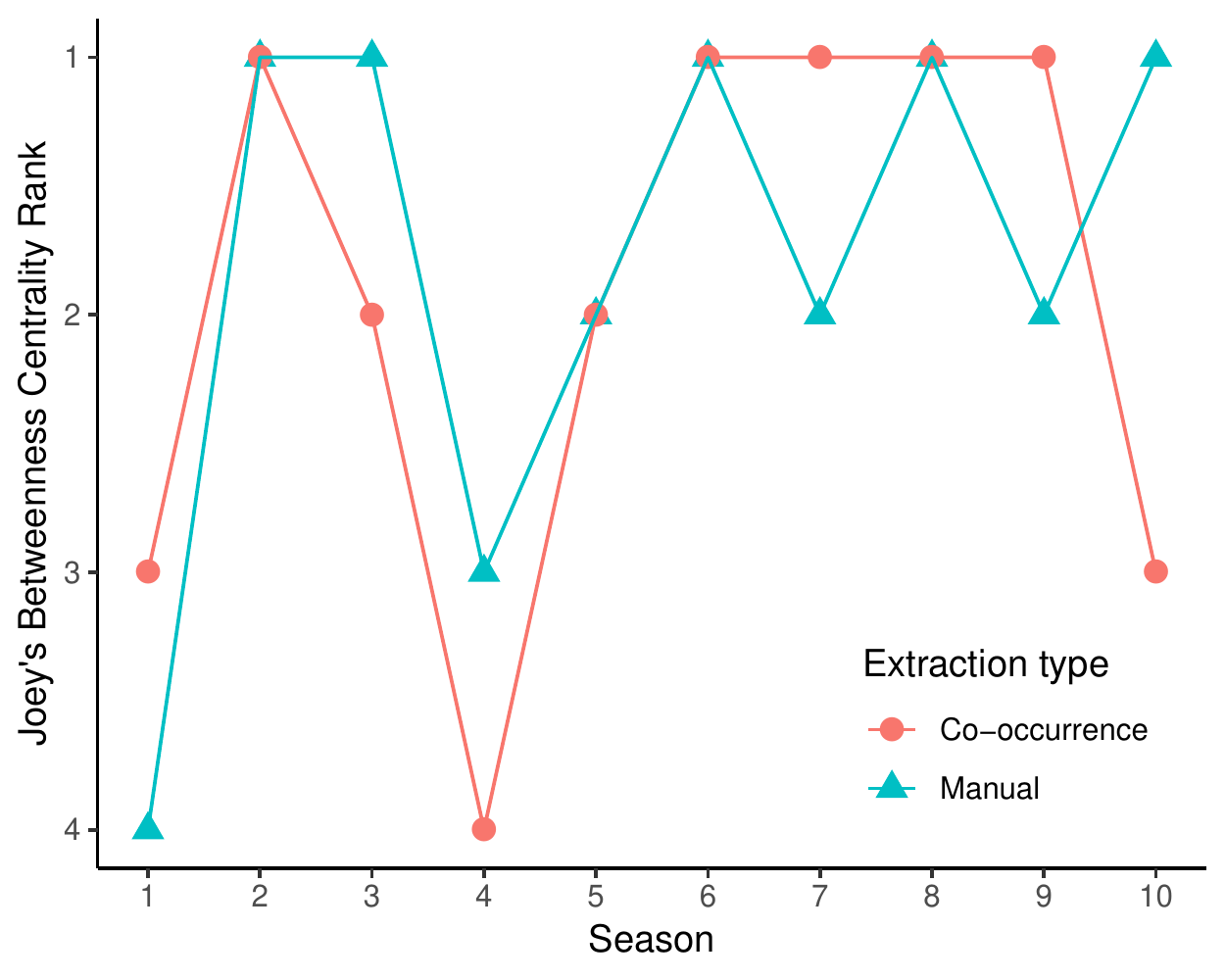}
\caption{The betweenness centrality ranks of Joey over the 10 seasons of \textit{Friends} for the \textbf{manual} and \textbf{co-occurrence network} datasets. Joey's betweenness centrality rank is very similar in every season, with a maximum difference of 2 in Season 10.}
\label{fig:eigenrank}
\end{figure}

We see a similar pattern in the data. \autoref{fig:eigenrank} shows the rankings of betweenness centralities of Joey in the real data for the \textbf{manual} and \textbf{co-occurrence} season networks. Joey has the highest or second highest betweenness ranking in every season except Season 1 and Season 4 (and Season 10 in the \textbf{co-occurrence network}). The rankings of betweenness centrality for the other core characters are in the supplementary information. While the exact value of the centrality may change in the different datasets, the ranking of the character is similar, so the analysis of character importance would be similar also.

The local clustering coefficient however is highly variable in the simulated \textbf{co-occurrence} and \textbf{NLP networks}. The correlations between the simulated \textbf{manual networks} and observation networks are moderate and positive on average, but range from -0.93 to 1. The large range of correlations show that we should not trust automatically extracted networks when looking at local clustering coefficients.

\subsection{Relationship metrics}

\begin{figure}[h]
\includegraphics[width = 0.9\textwidth]{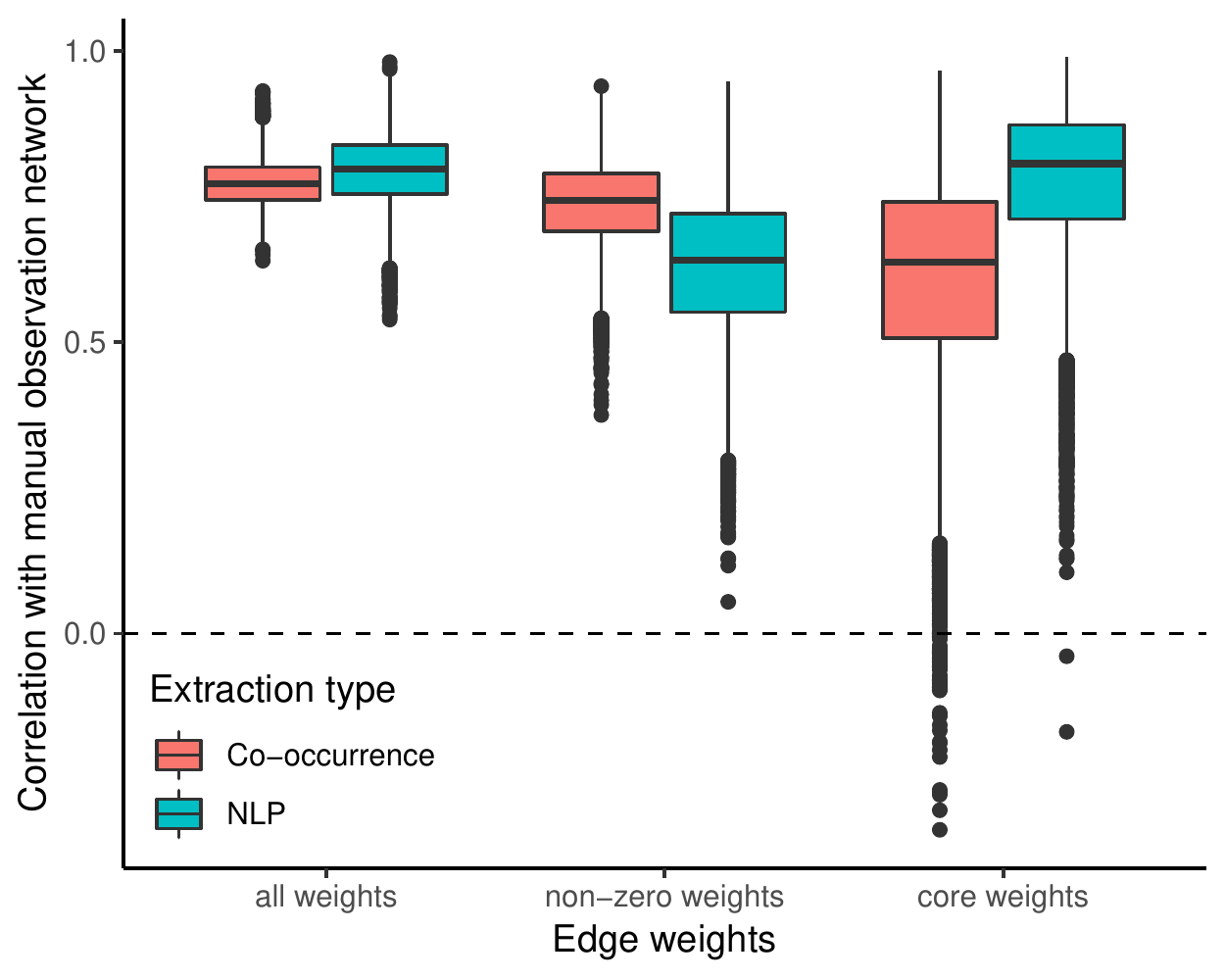}
\caption{Box-plots of the edge weight correlations between the simulated \textbf{manual network} and simulated \textbf{co-occurrence} and \textbf{NLP networks} from the 10000 simulations. The correlation is higher for both random networks when we include all edge weights. In both cases, the simulated \textbf{co-occurrence networks} are more reliable than the simulated \textbf{NLP networks}. If we only include edge weights between core characters there is a lot of variability in the correlations between the networks for both observation types.}
\label{fig:edge}
\end{figure}

Similarly to character metrics, we use edge weights to investigate the importance of relationships between characters. We correlate three sets of edge weights to asses the accuracy of different types of relationship analyses:
\begin{enumerate}
	\item All weights - including zero weights where there is no edge,
	\item Non-zero weights - between characters that interact at least once in at least one of the networks, and
	\item Core weights - between core characters, as these are usually the relationships we are most interested in.
\end{enumerate}
\autoref{fig:edge} shows the correlations between edge weights for the simulated \textbf{co-occurrence} and \textbf{manual networks} and simulated \textbf{NLP} and \textbf{manual networks}. Again, we use Spearman's correlation coefficient because we are interested in rank orderings rather than actual values.

%FIGURE fig:edge

The correlation between all edge weights in the simulated \textbf{manual} and \textbf{co-occurrence} networks are high with little variance. There is more spread in the correlation between edge weights in the simulated \textbf{NLP} and the \textbf{manual networks}. Correlations decrease when we exclude edges with zero weights in both networks.

But the non-zero edge weight correlations are still high for the simulated \textbf{co-occurrence networks}. This indicates that while we frequently get the correct set of interactions, the weights of those that interact are less accurate.

If we only compare the edges between the six core characters we find the correlations for both random observation networks vary greatly. Our results show the edge weight rankings in simulated \textbf{co-occurrence networks} are more reliable than in simulated \textbf{NLP networks} if we are interested in all characters. However, if we only analyse the relationships between core characters, the \textbf{NLP networks} are more reliable. This is because the core characters are frequently in scenes together but do not necessarily interact. This makes inferring the relative strengths of each relationship difficult when we only observe who is in the scene (\ie\ from the \textbf{co-occurrence network}), but \textbf{NLP networks} misdirect each interaction with the same probability, so edge weights between core characters are equally likely to be changed.

\subsection{Social networks in \textit{Friends}} \label{sec:lessfriendly}

Simulation results are very useful here because we know the true network, but the end goal is to apply these to real data. Here we compare metrics from the real datasets to analyse the social networks in \textit{Friends}. Visualisations of the networks and metrics are available at \url{friends-network.shinyapps.io/ingenuity_app/}. Here we focus on one finding, namely that the core \textit{Friends} get less ``friendly" over the ten seasons.

\begin{figure}[h]
\includegraphics[width = 0.9\textwidth]{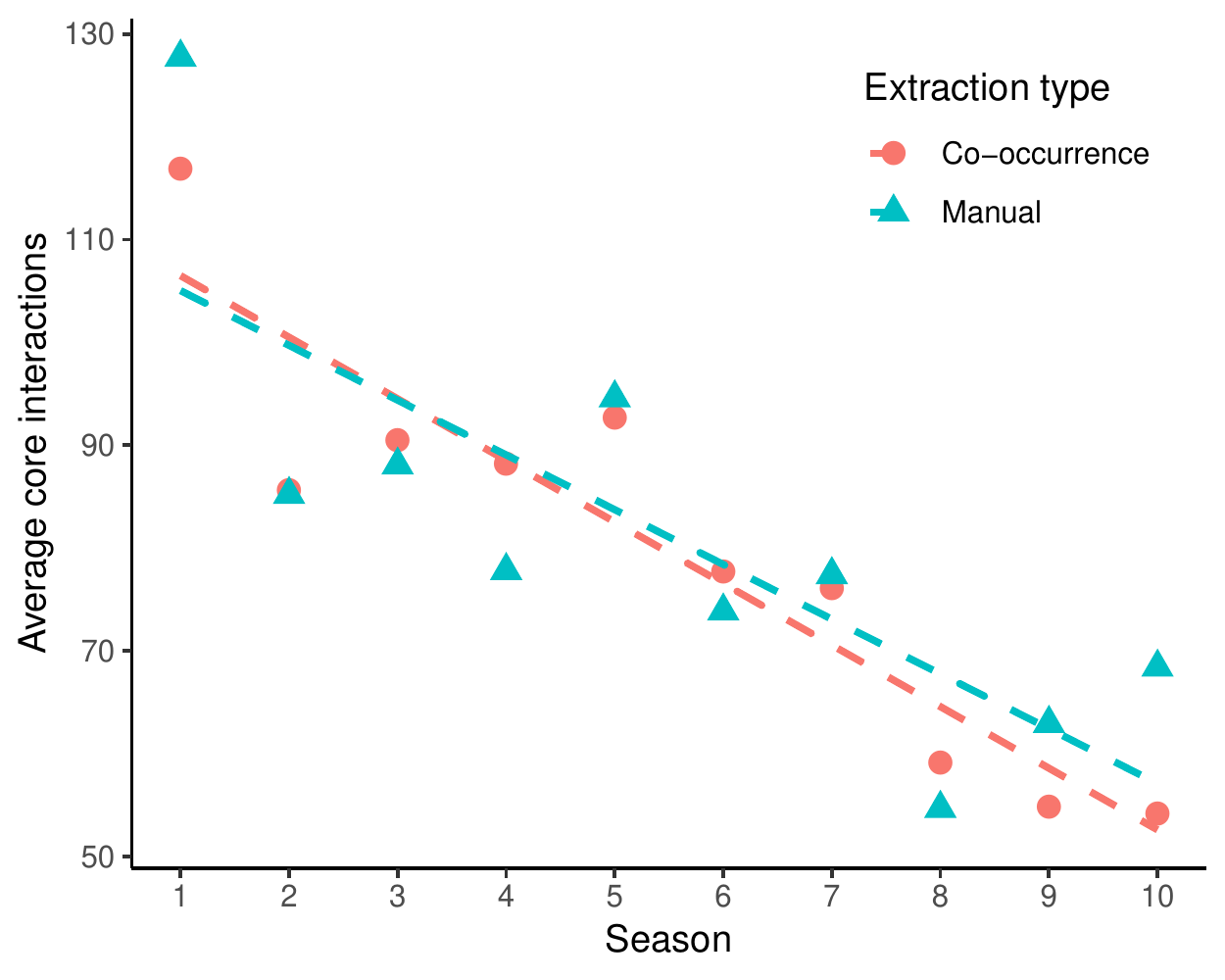}
\caption{Scatterplot of the average number of interactions between pairs of core characters in each season of \textit{Friends} from the co-occurrence and manual datasets. The dashed lines represent the line of best fit for the co-occurrence and manual network average core interactions in red and blue respectively. We see that although the individual datapoints vary, the trend is preserved.}
\label{fig:lessfriendly}
\end{figure}

\autoref{fig:lessfriendly} shows a scatterplot of the average number of interactions between pairs of core characters in each season of \textit{Friends} as inferred from the co-occurrence dataset and the manual dataset. We see that as the series develops, the core characters interact with each other less, \ie\ the \textit{Friends} get less friendly. This result is consistent between both datasets, with only slight variations in the slope and variance of the line of best fit.

 A possible explanation for this trend is that at the beginning of the series, the relationships between core characters need to be established, so they need to interact more. As the series develops, the interactions between core characters become repetitive, so these characters interact with each other less extras become more important. An interesting direction for future research is whether similar trends exist in other television series.

%FIGURE fig:lessfriendly1

%%%%%%%%%%%%%%%%
%% Conclusion %%
%%
\section{Conclusion}

The high correlation between the metrics of the manually extracted networks and the automatically extracted networks suggest that for most narrative analyses we can extract the social network automatically and achieve similar results to the more time consuming manual extraction. We should, however, keep in mind that this introduces some errors. For most metrics these errors will have minimal effects on the \emph{comparison} of global metrics over time and the importance of characters. A small set of metrics related to clustering are not reliable in the automatically extracted networks.

Here we only investigated the effect of the extraction method on the television show \(Friends\). With these comparison methods in place, one could check for consistent results in the other television shows and other types of narratives such as films and novels. 

The core group of characters in \textit{Friends} is intrinsic to the series, but extending the work to other narratives means we have to identify core characters in those narratives. A Stochastic Block Model \cite{weighted-sbm} could be used here to automatically identify the core group of characters. It is interesting to consider how the extraction approach might bias this identification, as some approaches to find core characters might be perturbed by distortion in clustering.

\section{Supplementary Information}

\begin{figure}[h]
\includegraphics[width = 0.95\textwidth]{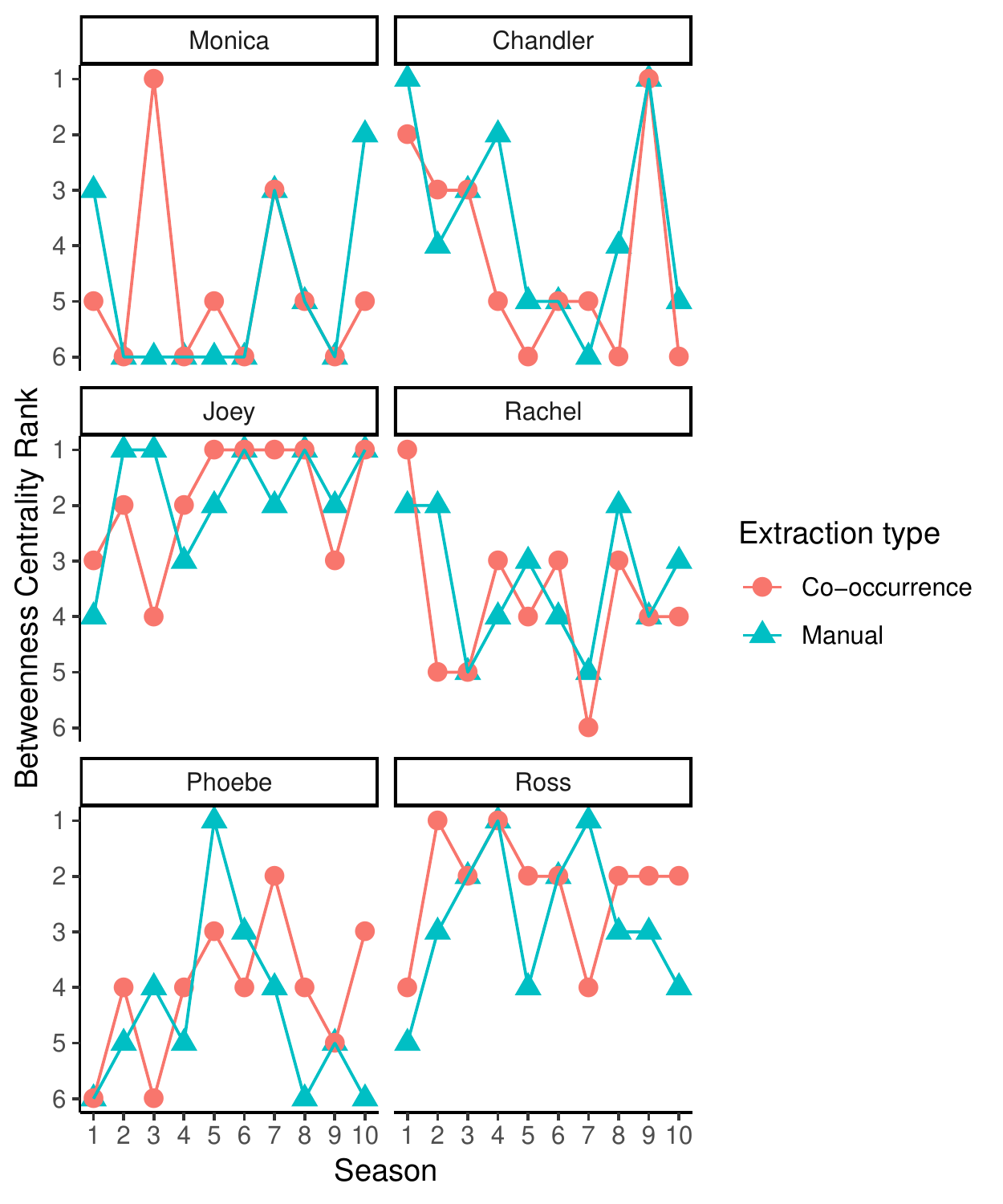}
\caption{Betweenness centrality ranks of each core character over every season for the co-occurrence and manual datasets.}
\label{fig:betweennessrank}
\end{figure}

\autoref{fig:betweennessrank} shows the betweenness centrality ranks for all core characters, analogous to \autoref{fig:eigenrank} in the main text. Although there are minor differences in many of the ranks from the co-occurrence networks compared to the manual networks, large differences are rare. Therefore analysis of characters using betweenness centrality is not greatly affected by the network extraction method. For example, the spike in Chandler's betweenness centrality rank in Season 9 is present in both datasets, as well as Joey's constantly high betweenness rank.

%%%%%%%%%%%%%%%%%%%%%%%%%%%%%%%%%%%%%%%%%%%%%%
%%                                          %%
%% Backmatter begins here                   %%
%%                                          %%
%%%%%%%%%%%%%%%%%%%%%%%%%%%%%%%%%%%%%%%%%%%%%%

\begin{backmatter}

\section*{Acknowledgements}
We acknowledge A. Bazzan for collecting and sharing her data with us and Australian Research Council Centre of Excellence for Mathematical and Statistical Frontiers (ACEMS) for funding the research.

\section*{List of abbreviations}
NLP, Natural Language Processing; sitcom, situation comedy.

\section*{Author's contributions}
ME collected and analysed the data; ME wrote the manuscript; LM, MR and JT drafted the manuscript. All authors read and approved the final manuscript.

\section*{Competing interests}
  The authors declare that they have no competing interests.

%%%%%%%%%%%%%%%%%%%%%%%%%%%%%%%%%%%%%%%%%%%%%%%%%%%%%%%%%%%%%
%%                  The Bibliography                       %%
%%                                                         %%
%%  Bmc_mathpys.bst  will be used to                       %%
%%  create a .BBL file for submission.                     %%
%%  After submission of the .TEX file,                     %%
%%  you will be prompted to submit your .BBL file.         %%
%%                                                         %%
%%                                                         %%
%%  Note that the displayed Bibliography will not          %%
%%  necessarily be rendered by Latex exactly as specified  %%
%%  in the online Instructions for Authors.                %%
%%                                                         %%
%%%%%%%%%%%%%%%%%%%%%%%%%%%%%%%%%%%%%%%%%%%%%%%%%%%%%%%%%%%%%

% useful resource: https://faculty.math.illinois.edu/~hildebr/tex/bibliographies0.html

% if your bibliography is in bibtex format, use those commands:
\bibliographystyle{bmc-mathphys} % Style BST file (bmc-mathphys, vancouver, spbasic).
\bibliography{../thesis}      % Bibliography file (usually '*.bib' )

\end{backmatter}
\end{document}